\documentclass[final,12pt,times]{elsarticle}

\usepackage{silence}
\usepackage{graphicx}

\usepackage{dcolumn}
\usepackage{braket}
\usepackage{amsmath}
\usepackage{amssymb}
\usepackage{bbm}
\usepackage{stackengine}
\usepackage{algorithm}
\usepackage{algpseudocode}

\usepackage{texlogos}

\usepackage{minted}

\usepackage[hidelinks]{hyperref}

\journal{Computer Physics Communications}

\makeatletter
\newenvironment{breakablealgorithm}
  {
   \begin{center}
     \refstepcounter{algorithm}
     \hrule height.8pt depth0pt \kern2pt
     \renewcommand{\caption}[2][\relax]{
       {\raggedright\textbf{\ALG@name~\thealgorithm} ##2\par}%
       \ifx\relax##1\relax 
         \addcontentsline{loa}{algorithm}{\protect\numberline{\thealgorithm}##2}%
       \else 
         \addcontentsline{loa}{algorithm}{\protect\numberline{\thealgorithm}##1}%
       \fi
       \kern2pt\hrule\kern2pt
     }
  }{
     \kern2pt\hrule\relax
   \end{center}
  }
\makeatother

\begin{document}

\begin{frontmatter}

\title{\textbf{Optimizing and benchmarking the computation of the permanent of general matrices}}

\date{\today}

\author[a]{Cassandra Masschelein \corref{first}}
\author[b]{Michelle Richer \corref{first}}
\author[a]{Paul W.\ Ayers \corref{author}}
\cortext[first]{Co-first authors.}
\cortext[author]{Corresponding author: \href{mailto:ayers@mcmaster.ca}{ayers@mcmaster.ca}}              
\address[a]{Department of Chemistry \& Chemical Biology, McMaster University, 1280 Main St.\ West, Hamilton, Ontario, L8S 4M1, Canada}
\address[b]{Department of Mathematics and Statistics, University of Ottawa, 75 Laurier Ave E, Ottawa, Ontario, K1N 6N5, Canada}

\begin{abstract}
Evaluating the permanent of a matrix is a fundamental computation that emerges in many domains, including traditional fields like computational complexity theory, graph theory, many-body quantum theory and emerging disciplines like machine learning and quantum computing. While conceptually simple, evaluating the permanent is extremely challenging: no polynomial-time algorithm is available (unless $\textsc{P} = \textsc{NP}$). To the best of our knowledge there is no publicly available software that automatically uses the most efficient algorithm for computing the permanent. In this work we designed, developed, and investigated the performance of our software package which evaluates the permanent of an arbitrary rectangular matrix, supporting three algorithms generally regarded as the fastest while giving the exact solution (the straightforward combinatoric algorithm, the Ryser algorithm, and the Glynn algorithm) and, optionally, automatically switching to the optimal algorithm based on the type and dimensionality of the input matrix. To do this, we developed an extension of the Glynn algorithm to rectangular matrices. Our free and open-source software package is distributed via Github, at \url{https://github.com/theochem/matrix-permanent}.
\end{abstract}


\end{frontmatter}

\clearpage

{\noindent\bf{Program Summary}}

\begin{small}
\noindent
{\em Program Title: } \texttt{matrix-permanent} \\
{\em Program file doi: } \url{https://github.com/theochem/matrix-permanent} \\
{\em Licensing provisions: } GNU General Public License v3.0 \\
{\em Programming language: } \cpluspluslogo{}, Python \\
{\em Supplementary material: Summary of Implemented Permanent Algorithms} \\
{\em Nature of problem: }
The permanent is a scalar-valued function of a matrix that is similar to the determinant but, because it is a sum over unsigned permutations, it has different mathematical properties. In particular, evaluating the permanent of a matrix has non-polynomial computational complexity \cite{minc1984permanents,valiant1979complexity}. The permanent arises in applied math (especially combinatorics and graph theory) and in adjacent fields of physics and chemistry. \\
{\em Solution method:}
The \texttt{matrix-permanent} library implements the most efficient algorithms for computing the permanents of general matrices, as the computational efficiency of the algorithm changes with dimension and density. The library's automatic tuning capabilities allow the most efficient algorithm for a matrix of some given dimensions to be chosen automatically. The library supports a diverse range of matrix types, including real, complex, binary, sparse, and dense matrices.\\
{\em Additional comments including restrictions and unusual features:}
The matrix dimensions where each algorithm is the most efficient can be determined automatically at compile-time, generating a function that always chooses the optimal algorithm for a given matrix, customized to the machine executing it.  \\
\\


\end{small}


\section{Introduction}

\subsection{Background}

In linear algebra, the determinant and permanent are both special cases of the immanant function of general square matrices, $\text{Imm} : \mathcal{M}_n(\mathbbm{F}) \rightarrow \mathbbm{F}$,
\begin{equation} \label{eq:immanant_definition}
    \text{Imm}(A) = \sum_{\sigma \in S_n}{{\chi_\lambda}(\sigma)\prod_{i=1}^n{a_{i \sigma(i)}}},
\end{equation}
where $\lambda$ is a partition of $n$ and $\chi_\lambda$ is the corresponding character of the symmetric group $S_n$. If $\chi_\lambda$ is the trivial (identity) character $1$, then Eq. \ref{eq:immanant_definition} is the \textit{permanent},
\begin{equation} \label{eq:permanent_definition}
    \text{per}(A) = \sum_{\sigma \in S_n}{\prod_{i=1}^n{a_{i \sigma(i)}}},
\end{equation}
and if $\chi_\lambda$ is the alternating character $\text{sgn}$, then Eq. \ref{eq:immanant_definition} is the \textit{determinant},
\begin{equation} \label{eq:determinant_definition}
    \text{det}(A) = \sum_{\sigma \in S_n}{\text{sgn}(\sigma)\prod_{i=1}^n{a_{i \sigma(i)}}}.
\end{equation} Despite their similar definitions, these functions have vastly different computational properties. While it is well known that the determinant can be computed as efficiently as a matrix multiplication ($\mathcal{O}(n^a)$) via matrix decompositions like Gaussian elimination ($a = 3$) or via Strassen's method ($a \approx 2.479$) \cite{bunch1974triangular,strassen1969gaussian}, these methods are not applicable to the computation of permanents because the permanent is not a multilinear form. Indeed, Valiant proved in 1979 that the computation of the permanent is in the complexity class \textsc{\#P-complete} \cite{valiant1979complexity}, and therefore no efficient (polynomial time) algorithm exists to compute it (presuming that $\textsc{P} \neq \textsc{NP}$).

While both the determinant and the permanent have a clear geometric interpretation, the permanent is also a \emph{combinatoric} object and is related to perfect matching in graph theory. Also unlike the determinant, it is meaningful to compute the permanents of rectangular matrices, using the more general signature $\text{per} : \mathcal{M}_{mn}(\mathbbm{F}) \rightarrow \mathbbm{F}$, since the definition of the rectangular permanent,
\begin{equation} \label{eq:rectangular_permanent_definition}
    \begingroup
    \renewcommand*{\arraystretch}{1.6}
    \text{per}(A) = \left\{\begin{matrix} 
        \,\,\sum\limits_{\sigma \in P_{n,m}}{\prod\limits_{i=1}^{m}{a_{i \sigma(i)}}}\,\, & m \leq n \\
        \text{per}(A^T) & m > n 
    \end{matrix}\right.
    \endgroup
\end{equation}
where $P_{n,m}$ is the $m$-permutation set of $\{1, \dots, n\}$, still has clear combinatoric and graph-theoretic interpretations \cite{valiant1979complexity, gurvits2014bounds, guichard2017introduction, petterson2009exponential, dufosse2022scaling, david2013introduction}.

\subsection{Computing the permanent}

Computing the permanent using its definition (Eqs.\ \ref{eq:permanent_definition}, \ref{eq:rectangular_permanent_definition}) has computational complexity $\mathcal{O}(m\,n!/(n-m)!)$, which in the square case reduces to $\mathcal{O}(n\, \cdot n!)$. Below we discuss the current most efficient algorithms for general matrices, which have better scaling than this.

\paragraph{Inclusion-exclusion principle approach \label{ryser_algo}}
Ryser proposed an algorithm for computing the permanent based on the inclusion-exclusion principle for computing the cardinality of set unions, e.g., $|A \cup B| = |A| + |B| - |A \cap B|$. For a matrix $A \in \mathcal{M}_{mn}(\mathbbm{F})$, we define $\mathcal{A}_r$ as the set of matrices obtained by replacing $r$ columns of $A$ with columns of zeroes (or by ``deleting'' the columns), and $R(A)=\prod_{i=1}^m{\sum_{j=1}^n{a_{ij}}}$ as the product of row-sums of $A$. We can then use the inclusion-exclusion principle to reformulate Eq.\ \ref{eq:rectangular_permanent_definition} as
\begin{equation}
\text{per}(A) = \sum_{k=0}^{m-1}{(-1)}^k{\sum_{A'_k \in \mathcal{A}_k}{R(A'_k)}}
\end{equation}

This gives the general Ryser formula,
\cite{ryser1963combinatorial}
\begin{equation} \label{eq:ryser_rectangle}
    \text{per}(A) = \sum_{k=0}^{m-1}{{(-1)}^k \sum_{\sigma \in P_{n,m-k}}{  \binom{\,n - m + k\,}{k} \prod_{i=1}^{m}{\sum_{j=1}^{m-k}{a_{i \sigma(j)}}}}},
\end{equation}
which, for square matrices, reduces to 
\begin{equation} \label{eq:ryser_square}
    \text{per}(A) = {(-1)}^n \sum_{k=1}^n{{(-1)}^k \sum_{\sigma \in P_{n,k}}{\prod_{i=1}^n{\sum_{j=1}^k{a_{i \sigma(j)}}}}}.
\end{equation}
This algorithm scales as $\mathcal{O}(2^n \cdot  n)$, assuming that the permutations are iterated over in minimal change order (e.g., by the Steinhaus-Johnson-Trotter algorithm \cite{trotter1962algorithm,knuth2014art,arndt2010matters}, which our implementation uses). The base implementation for the square and rectangular cases is provided in~\ref{app:algorithms} (Algorithms~\ref{app:alg:ryser_square}–\ref{app:alg:ryser_rectangular}).

\paragraph{Invariant theory approach \label{glynn_algo}}
Glynn proposed an alternative approach where the polarization identity for symmetric tensors is used to deduce the following expression, valid for square matrices,~\cite{glynn2010permanent}
\begin{equation} \label{eq:glynn}
    \text{per}(A) = \frac{1}{2^{n-1}} \sum_{\delta \in \{\pm1\}^n}{\Big(\prod_{k=1}^n{\delta_k \Big) \prod_{j=1}^n{\sum_{i=1}^n{\delta_i a_{ij}}}}} .
\end{equation}
To extend this to rectangular matrices, we add additional rows of 1's \cite{bebiano1982evaluation},
\begin{subequations} \label{eq:glynn_filled_matrix}
\begin{align}
    A &\in \mathcal{M}_{mn}(\mathbbm{F}) = (a_{ij})\,,~m<n \\
    {A'} &\in \mathcal{M}_{nn}(\mathbbm{F}) = 
    \left[\phantom{\begin{matrix}a_0\\ \ddots\\a_0\\b_0\\ \ddots\\b_0 \end{matrix}}
    \right.\hspace{-1.3em}
    \underbrace{\begin{matrix}
        a_{11} & \cdots & a_{1n} \\
        \vdots & \ddots & \vdots \\
        a_{m1} & \cdots & a_{mn} \\
        1 & \cdots & 1 \\
        \vdots & \ddots & \vdots \\
        1 & \cdots & 1
    \end{matrix}}_{\displaystyle{n}}
    \hspace{-1.3em}
    \left.\phantom{\begin{matrix}a_0\\ \ddots\\a_0\\b_0\\ \ddots\\b_0 \end{matrix}}\right]\hspace{-1em}
    \begin{tabular}{l}
    $~\left.\lefteqn{\phantom{\begin{matrix} a_0\\ \ddots\\ a_0\ \end{matrix}}}\right\}m$\\
    $~\left.\lefteqn{\phantom{\begin{matrix} b_0\\ \ddots\\ b_0\ \end{matrix}}} \right\}n - m$
    \end{tabular} 
\end{align}
\end{subequations}
This gives the working formula,
\begin{subequations} \label{eq:glynn_rectangle}
\begin{align}
    \text{per}(A) &= \frac{1}{(n-m)!}\text{per}(A') \\
    &= \frac{1}{2^{n-1}(n-m)!} \sum_{\delta \in \{\pm1\}^n}{\Big(\prod_{k=1}^n{\delta_k \Big) \prod_{j=1}^n{\Big(\sum_{i=1}^m{\delta_i a_{ij}} + \sum_{i=m+1}^n{\delta_i\Big)}}}}.
\end{align}
\end{subequations}
This algorithm also scales as $\mathcal{O}(2^n \cdot n)$, assuming that the permutations are iterated over in Gray code. For all implementations of the Glynn algorithm, we have used this variant. The base implementation for the square and rectangular cases is provided in~\ref{app:algorithms} (Algorithms~\ref{app:alg:glynn_square}–\ref{app:alg:glynn_rectangular}).

\subsection{Applications}

The (rectangular) permanent is a fundamental function in several areas of research, making the development of efficient algorithms for its computation an active area of research \cite{jerrum2004polynomial, richer2024geminalgraph, limacherNewWavefunctionHierarchy2016}. Some important applications of the permanent follow.

\paragraph{Graph theory}
The number of perfect matchings of a bipartite graph $G = (U, V, E)$ with disjoint sets of vertices $U$ and $V$ with respective cardinalities $m$ and $n$ can be found by counting the number of perfect matchings of the graph, which is equivalent to computing the permanent of the adjacency matrix $A$ \cite{minc1984permanents,david2013introduction,dufosse2022scaling}. For a simple adjacency matrix $A \in \mathcal{M}_{mn}({\{0, 1\}})$, this gives the number of perfect matchings, while for the adjacency matrix of a weight graph $A \in \mathcal{M}_{mn}(\mathbbm{R})$ this gives the sum of weights of the perfect matchings.

\paragraph{Quantum many-body problems} 
In quantum mechanics, the state of a system is described by its wavefunction, which is a vector in Hilbert space, $\ket{\Psi}$. The conjugate transpose of the vector is denoted $\bra{\Psi}$. Thus the overlap between two wavefunctions can be denoted $\braket{\Phi|\Psi}$ and the projection of $\ket{\Psi}$ onto the direction defined by $\ket{\Phi}$ is $\ket{\Phi}\braket{\Phi|\Psi}$. Observable quantum-mechanical properties correspond to Hermitian operators, and the value of the observable is determined by $\braket{\Psi|\hat{H}\Psi} = \braket{\hat{H}\Psi|\Psi} = \braket{\Psi|\hat{H}|\Psi}$. 

Mathematically, a system of $n$ hard-core bosons, each of which can occupy any of $N$ single-boson states, can be represented as a quantum superposition (linear combination) of all possible ways to occupy these states, 
\begin{equation}\label{eq:DOCI}
   \ket{\Psi} =  \sum_{\Set{m_j \in \{0,1\}|n = \sum_{j=1}^N m_j}} c_{m_1 m_2 \ldots m_N} \left(b_1^\dagger \right)^{m_1} \left(b_2^\dagger \right)^{m_2}\cdots \left(b_N^\dagger \right)^{m_N} \ket{\emptyset}
\end{equation}
where $\ket{\emptyset}$ is the (physical) vacuum (all states are empty) and the operator $\left(b_j^\dagger\right)^{m_j}$ creates a boson in the $j$-th state if $m_j=1$ and does nothing (multiplication by 1) if $m_j = 0$. 
The occupation-number vectors form an orthogonal and normalized basis for the Hilbert space of wavefunctions, and it is convenient to represent occupation-number vectors as bitstrings, e.g. $\ket{\mathbf{m}} = \ket{m_1 m_2 \ldots m_N}$. The wavefunction (\ref{eq:DOCI}) appears in electronic structure theory (where each $b_j^\dagger$ corresponds to the creation of an electron pair), quantum computing (where $b_j^\dagger$ is the operator that converts a 0-qubit to a 1-qubit), and spin physics (where $b_j^\dagger$ flips a down-spin particle to an up-spin particle) \cite{bytautasSeniorityOrbitalSymmetry2011c,chuikoModelHamiltonian2024}.

A compact, but superficially approximate, mean-field parameterization of the wavefunction, $\ket{\Psi}$, is obtained by taking a linear transformation of the boson-creation operators,\cite{johnsonSizeconsistentApproachStrongly2013,richerGraphicalApproachInterpreting2025,johnsonStrategiesExtendingGeminalbased2017b,kim2021flexible,apig_first,
limacherNewMeanfieldMethod2013}
\begin{equation}
    B_i^\dagger = \sum_{j=1}^N c_{ij} b_j^\dagger
\end{equation}
and then constructing a symmetric product of these boson states (SBP), 
\begin{equation}\label{eq:SBP}
    \ket{\Psi_{\text{SBP}}} = \prod_{i=1}^n B_i^\dagger \ket{\emptyset}.
\end{equation}
We would like to be able to evaluate the coefficients in Eq. (\ref{eq:DOCI}) for the SBP wavefunction. For a given $N$-boson state, $\ket{\mathbf{m}}$ with $m_j=1$, the creation of the $m_j$-th boson could be associated with any of the $N$ boson creation operators, $B_i^\dagger$, introducing a multiplicative factor of $c_{ij}$. Summing over all possible ways to create the occupations in $\ket{\mathbf{m}}$ is equivalent to evaluating the permanent of a $N$-by-$N$ matrix. Specifically, the $N$ columns of the matrix where $m_j=1$ are filled in with the elements of $c_{ij}$. Numerically, this corresponds to multiplying the $N$-by-$n$ matrix with elements $c_{ij}$ by a $n$-by-$N$ matrix that is entirely zero, but except there is a 1 in the column $j$ if $m_j$ is the $j$-th nonzero entry in $\ket{\mathbf{m}}$. 
\begin{equation}
    p_{ij} = 0 \quad \text{unless} \quad m_j = 1 \quad \text{and} \quad j = 1 + \sum_{k=1}^{i-1} m_k
\end{equation}
This is a generalized permutation matrix, where the row-sums are zero or one, but the column sums are one, and corresponds to a way to select $N$ objects from $n$ choices. One can then write
\begin{equation}\label{eq:SBGpermanent}
    c_{\mathbf{m}} = \braket{\mathbf{m}|\Psi_{SBP}} = \text{per}(\mathbf{C}\mathbf{P}).
\end{equation}

To support this use case, \texttt{ matrix permanent} was integrated with the \texttt{PyCI} package for solving the quantum-many body problem \cite{richer2024pyci}.

\paragraph{Permanental Point Processes}
Determinantal point processes are associated with distributions of fermions in space and are commonly used to generate samples of points where clustering is less likely to occur than with random (Poisson) processes \cite{kuleszaDeterminantalPointProcesses2012}. Permanental point processes are associated with distributions of bosons in space and are used to generate samples of points where clustering is prevalent \cite{baakeDiffractionTheoryPoint2015,
eisenbaumPermanentalProcesses2009,
hultgrenPermanentalPointProcesses2019,
jahangiriPointProcessesGaussian2020,
kimFastBayesianEstimation2022,
mccullaghPermanentalProcess2006}.

To understand where permanental point processes arise, recall that the elements of the one-boson reduced density matrix (1DM) for an $n$-boson system can be evaluated as:
\begin{equation}
    \gamma_{ij} = \Braket{\Psi|b_{i}^\dagger b_j|\Psi}.
\end{equation}

Without any information about the higher-order reduced density matrices, the closest one can come to estimating the complete $n$-boson density matrix is defined by the permanent,
\begin{equation}
    \ket{\Psi}\bra{\Psi} \approx \text{per} \begin{bmatrix}
\gamma_{p_1 q_1} & \gamma_{p_1 q_2} & \cdots &\gamma_{p_1 q_n}\\
\gamma_{p_2 q_1} & \gamma_{p_2 q_2} & \cdots &\gamma_{p_2 q_n}\\
\vdots & \vdots & \ddots &\vdots\\
\gamma_{p_n q_1} & \gamma_{p_n q_2} & \cdots &\gamma_{p_n q_n}
\end{bmatrix}.
\end{equation}
The correction to this approximation is given by the second-order cumulant \cite{kuboGeneralizedCumulantExpansion1962,
ziescheCumulantExpansionsReduced2000}.

Note that the 1DM is positive semidefinite by construction, and can be expressed as a kernel,
\begin{equation}
    \gamma(\mathbf{r},\mathbf{r}') = \sum_{i,j} \gamma_{ij} \phi_i(\mathbf{r}) \phi_j^*(\mathbf{r}')
\end{equation}
where $\phi_i(\mathbf{r}) = b_i^\dagger \ket{\emptyset}$ is a single-boson state. If one diagonalizes this matrix, one gets the normal ``kernel form'' that is used in point processes,\cite{mccullaghPermanentalProcess2006,kuleszaDeterminantalPointProcesses2012} 
\begin{equation}
    \gamma(\mathbf{r},\mathbf{r}') = \sum_{i} \lambda_i \chi_i(\mathbf{r}) \chi_i^*(\mathbf{r}')
\end{equation}
where $\lambda_i \ge 0$. One can then write the density matrix as a permanent,
\begin{equation}
    \Psi(\mathbf{r}_1,\mathbf{r}_2,\ldots,\mathbf{r}_n) \Psi^*(\mathbf{r}_1',\mathbf{r}_2',\ldots,\mathbf{r}_n') \approx \text{per} \begin{bmatrix}
\gamma(\mathbf{r}_1,\mathbf{r}_1') & \gamma(\mathbf{r}_1,\mathbf{r}_2') & \cdots &\gamma(\mathbf{r}_1,\mathbf{r}_n')\\
\gamma(\mathbf{r}_2,\mathbf{r}_1') & \gamma(\mathbf{r}_2,\mathbf{r}_2') & \cdots &\gamma(\mathbf{r}_2,\mathbf{r}_n')\\
\vdots & \vdots & \ddots &\vdots\\
\gamma(\mathbf{r}_n,\mathbf{r}_1') & \gamma(\mathbf{r}_n,\mathbf{r}_2') & \cdots &\gamma(\mathbf{r}_n,\mathbf{r}_n')
\end{bmatrix}.
\end{equation}

The probability of observing bosons at the points $(\mathbf{r}_1,\mathbf{r}_2,\ldots,\mathbf{r}_n)$ is given by the square-magnitude of the wavefunction,
\begin{equation}\label{eq:permpointprocess}
    p(\mathbf{r}_1,\mathbf{r}_2,\ldots,\mathbf{r}_n) \approx \text{per} \begin{bmatrix}
\gamma(\mathbf{r}_1,\mathbf{r}_1) & \gamma(\mathbf{r}_1,\mathbf{r}_2) & \cdots &\gamma(\mathbf{r}_1,\mathbf{r}_n)\\
\gamma(\mathbf{r}_2,\mathbf{r}_1) & \gamma(\mathbf{r}_2,\mathbf{r}_2) & \cdots &\gamma(\mathbf{r}_2,\mathbf{r}_n)\\
\vdots & \vdots & \ddots &\vdots\\
\gamma(\mathbf{r}_n,\mathbf{r}_1) & \gamma(\mathbf{r}_n,\mathbf{r}_2) & \cdots &\gamma(\mathbf{r}_n,\mathbf{r}_n)
\end{bmatrix}.
\end{equation}\

These equations are all exact for noninteracting (uncorrelated) bosons, and sampling with respect to Eq. \ref{eq:permpointprocess} is the permanental point process. 

\paragraph{Photonic Quantum Computers}
Photonic quantum computers are well adapted to evaluating the permanent of unitary matrices. While the $\ket{\Psi_{SBP}}$ wavefunction is not unitary (because $\mathbf{C}$ is not usually restricted to unitary transformations), one can add rows and columns to non-unitary matrices so that they become unitary \cite{levy2014dilation,ticozzi2017quantum,schlimgen2021quantum,suri2023two,buscemi2003physical,hu2020quantum,mazzola2024quantum,oh2024singular}. (This is called the unitary dilation of the operator \cite{schafferUnitaryDilationsContractions1955,sz.-nagyHarmonicAnalysisOperators2010}). This means that photonic quantum computers, if they had sufficient accuracy, could be used to evaluate the permanent of arbitrary matrices and, thereby, efficiently solve problems in $\# \textsc{P}$. One of the most important applications of algorithms for evaluating the matrix permanent is to perform (classical) simulations of photonic quantum computers.

\subsection{Approximate computation and special cases}

Although not the focus of this work, it is worth mentioning that the permanent can be computed (very) approximately far cheaper than the algorithms presented here can achieve; there are also structured matrices for which the permanent is cheaper to compute via special methods than by the general ones presented above.

\paragraph{Approximate computation} Algorithms exist to approximately compute the permanent of low-rank and positive semidefinite matrices, but the best class of algorithms for fully general matrices are Gurvits' randomized algorithms, which approximately compute the permanent of an $n$-by-$n$ matrix $A$ with time complexity $\mathcal{O}(n^2/\varepsilon^2)$ to within $\pm\varepsilon\|A\|^n$ \cite{gurvits2005complexity,aaronson2014generalizing}. This can often provide sufficient accuracy in cases where one samples permanents over a distribution, but not where high accuracy is required.

\paragraph{Low-rank matrices} Methods for computing the permanents of low-rank matrices (i.e., with repeated rows or columns) have been developed in the context of boson sampling. By starting with the Ryser algorithm, and taking into account the number of unique subsets of rows or columns, given the repetitions, the algorithm can be reformulated with a lower computational complexity \cite{shchesnovich2013asymptotic,clifford2024faster}. A similar method based on the Glynn algorithm also exists, with an improved constant prefactor \cite{chin2018generalized}. 

\paragraph{Cauchy matrices} The permanent of a Cauchy matrix, with elements $c_{ij} = {1/(x_i-y_j)}$, is easy to compute. Borchardt's theorem gives (originally for $m=n$, although this was trivially extended to rectangular matrices with $m<n$) \cite{borchardt1857bestimmung,carlitzIdentityCayley1960,han2000generalisation,han2000rectangular,chavezInductiveProofBorchardt2024}:
\begin{equation}\label{eq:caucy_permanent}
    \text{per}(C) = \frac{\text{det}(C \circ C)}{\text{det}(C)}
\end{equation}
Permanents of Cauchy matrices naturally appear in models for superconductivity and, more generally, electron pairing \cite{richardsonExactEigenstatesPairingforce1964,
richardsonRestrictedClassExact1963,johnsonBivariationalPrincipleAntisymmetrized2022,
johnsonSizeconsistentApproachStrongly2013,
tecmerAssessingAccuracyNew2014a,
johnsonRichardsonGaudinMeanfield2020,
johnsonSingleReferenceTreatment2023,
fecteau2022near,dukelskyColloquiumExactlySolvable2004,
dukelskyExactlySolvablePairing2007,
dukelskyExactlySolvablePairing2011b,
ortizExactlysolvableModelsDerived2005,
romboutsSolvingRichardsonEquations2004,faribaultDeterminantRepresentationsGaudin2012,
faribaultReducedDensityMatrices2022,
fecteauNearexactTreatmentSeniorityzero2022,
gaudinDiagonalisationClasseHamiltoniens1976,
gaudinDiagonalizationClassSpin1976,
gaudinSystemeDimensionFermions1967,
johnsonRichardsonGaudinStates2023,kim2023fanpy}.

\section{Library structure}

The matrix-permanent library consists of three parts: (a) a header-only \cpluspluslogo{} library implementing the permanent algorithms; (b) a program which can be run to generate parameters allowing dispatch to the most efficient algorithm based on the dimensions of the input matrix; and (c) a Python C extension module using the \cpluspluslogo{} library which allows the computation of permanents of NumPy arrays (\texttt{numpy.ndarray}) \cite{harris2020array} in Python.

\subsection{Automatic tuning of the library}



Baselines for automatic decision making are precomputed and stored in the default tuning file. When the program is compiled by a user they have the option to customize the tuning parameters to their machine. To do so the user simply needs to include the tuning flag when compiling the program for the first time by specifying \mintinline{bash}{make RUN_TUNING=1}. This will automatically re-generate the tuning file to be shipped with the \cpluspluslogo{} library and compiled into the Python C extension module.

\begin{figure}[htbp]
  \centering
  \includegraphics[width=\linewidth]
  {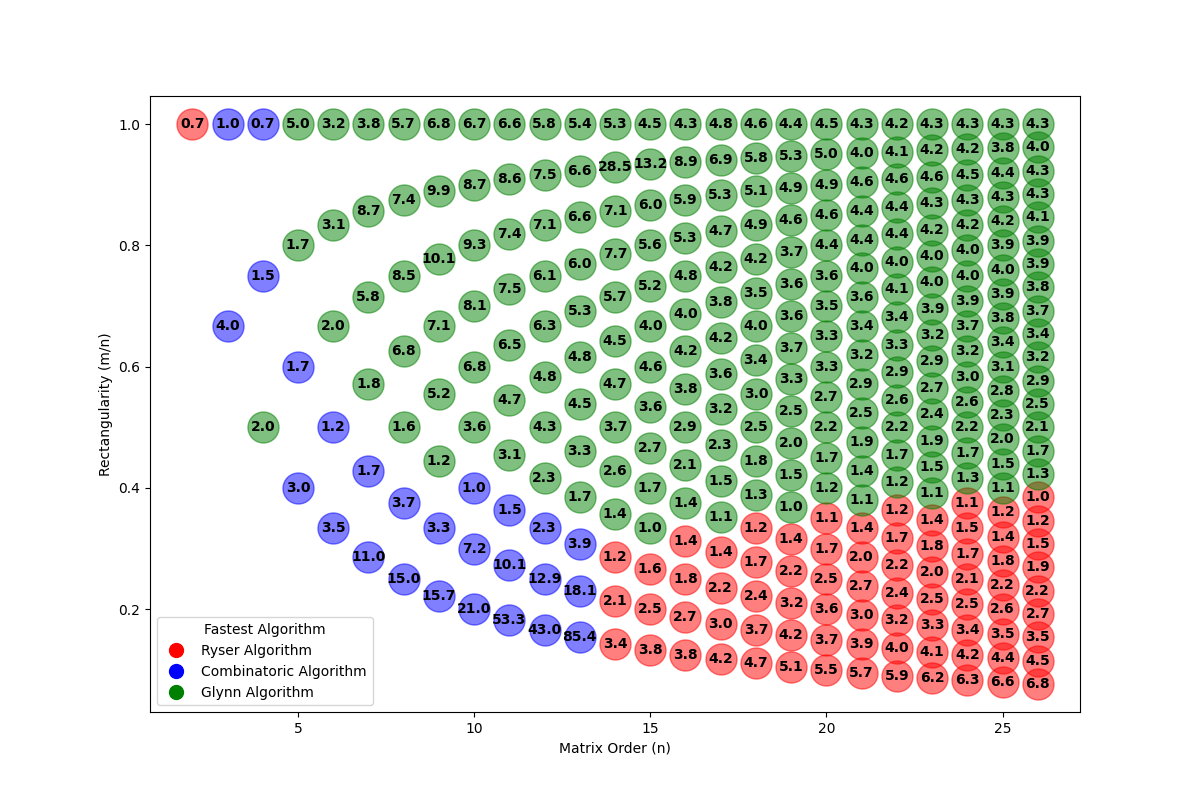}
  \caption{The fastest algorithm by matrix order, $n$, and the degree of rectangularity, $\tfrac{m}{n}$. The values reported indicate the performance (as a factor of execution time) that would be lost were the second-fastest algorithm used instead of the fastest algorithm.}
  \label{fig:fastest_scatter_nums}
\end{figure}

During our preliminary investigations, we concluded that the na\"ive combinatoric algorithm is infeasible for larger matrices. This finding, along with the linearly separable algorithm boundaries, allowed us to automate the tuning of the library by training two hard-margin support vector machines \cite{vapnik1995support, cortes1995support, awad2015support, boser1992training, weston1998multi, collobert2004links, crammer2001algorithmic, rosenblatt1957perceptron, rosenblatt1958perceptron, rosenblatt1961principles, freund1998large, block1962perceptron} with linear kernels. To allow this simplification, we automatically detect and hard-code the few cases where the Ryser algorithm consistently outperforms the na\"ive algorithm for small matrices and output them to the header file as a parameter. The resulting hyperplanes are then used to define the other parameters for the optimized algorithm swapping procedure. The procedure used to define the optimally algorithm switching is provided in Algorithm~\ref{app:alg:opt_permanent}. Note that near the decision boundaries, there is little performance penalty for choosing the second-best algorithm (see Figure \ref{fig:fastest_scatter_nums}); this justifies our decision to use a simple linear kernel. Also, for very small matrices, the computation is extremely fast, and choosing a suboptimal algorithm is unlikely to be detrimental.

\begin{figure}[htbp]
  \centering
  \includegraphics[width=\linewidth]{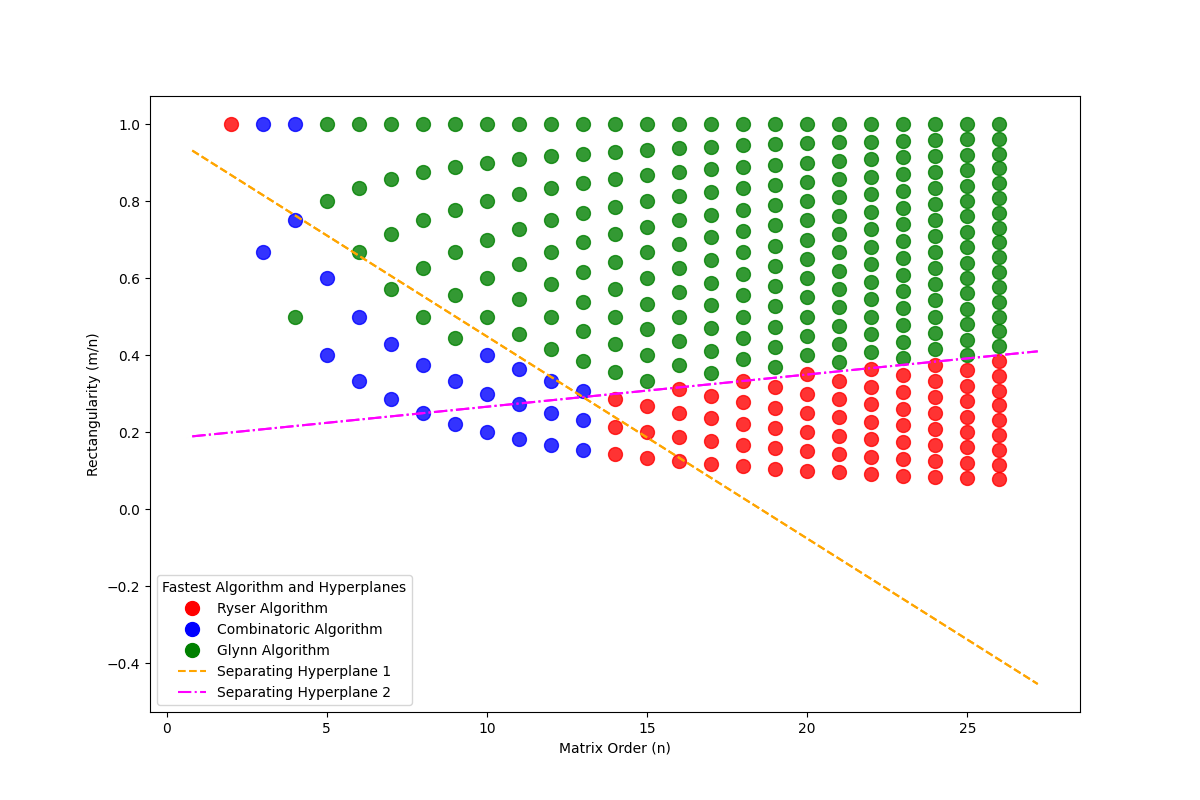}
  \caption{The space of fastest algorithms for computing the permanent of a matrix. The separating hyperplanes between the algorithms are depicted by the dotted lines. The hyperplanes intersect at matrix order $n = 13$ and rectangularity $m/n = 0.29$} 
  \label{fig:svm_hyperplanes}
\end{figure}

Tuning the algorithm results in two hyperplanes, separating the space into three regions; see Figure \ref{fig:svm_hyperplanes}. As expected, the combinatoric algorithm is best for very small matrices. For large matrices, the Glynn algorithm is normally preferable, but because treating rectangular matrices by augmentation with 1's (cf. Eq. (\ref{eq:glynn_filled_matrix})) is inefficient, the Ryser algorithm is more efficient for very rectangular matrices. 

\section{Usage}

\subsection{Installation}

The \texttt{matrix-permanent} library is hosted on GitHub, and can be installed via \texttt{pip}. Installation requires a \cpluspluslogo{} compiler, Python, and CMake. The header-only \cpluspluslogo{} library is located in the \texttt{include} sub-directory and can be used as-is or by including the repository as a CMake library and linking your target(s) to the \texttt{MatrixPermanent::headers} target. Or, to compile the Python extension module manually via CMake, set the CMake variable \texttt{PERMANENT\_PYTHON} to \texttt{ON}.
To install the Python extension module normally via \texttt{pip}, run:
\begin{minted}{bash}
git clone https://github.com/theochem/matrix-permanent
cd matrix-permanent
pip install .
\end{minted}
If you want to generate a machine-specific tuning header, set the CMake variable \texttt{PERMANENT\_TUNE} to \texttt{ON}, or preface the \texttt{pip} command with the corresponding environment variable like so:
\begin{minted}{bash}
PERMANENT_TUNE=ON pip install .
\end{minted}

\subsection{Using the \cpluspluslogo{} library}\label{sec:using_cpp_lib}

The matrix-permanent \cpluspluslogo{} library can be made available by including the header file:
\begin{minted}{cpp}
#include <permanent.h>
\end{minted}
The \cpluspluslogo{} library provides functions in the \texttt{permanent} namespace with the following signature, where the return type \texttt{result\_t<Type, IntType>} is either (a) if \texttt{IntType} is unspecified, type \texttt{double} or \texttt{std::complex<double>} depending on if \texttt{Type} is complex or (b) type \texttt{IntType} or \texttt{std::complex<IntType>} if \texttt{Type} is a (complex) integer type and \texttt{IntType} is a (complex) integer type. The simplest behaviour (when specifying just \texttt{Type}) is to always return a \texttt{double} or \texttt{std::complex<double>}, while \texttt{IntType} can be specified if the user wants to return an integer type when \texttt{Type} is also an integer type; \texttt{IntType} \emph{must} be specified in this case because the choice of integer return type must be made with consideration given to the contents of the input matrices and whether overflows or underflows are likely to occur.
\begin{minted}{cpp}
template<typename Type, typename IntType = void>
permanent::result_t<Type, IntType>
permanent::fn(const size_t m, const size_t n, const Type *ptr);
\end{minted}
The function name \texttt{fn} can be one of \{\texttt{combinatoric}, \texttt{glynn}, \texttt{ryser}, \texttt{opt}\}, which works for both square and rectangular matrices. Each of these names can also be given the suffix \texttt{\_square} or \texttt{\_rectangular}, e.g., \texttt{glynn\_square}, which directly dispatches the correct algorithm for the matrix shape. The \texttt{opt} functions use the tuning parameters to dispatch the most efficient algorithm for the input \texttt{m} and \texttt{n}. The pseudocode for the \texttt{opt} algorithm is given in~\ref{app:opt_algorithm}, Algorithm~\ref{app:alg:opt_permanent}.

\subsection{Using the Python C extension module}

The \texttt{permanent} C extension module can be directly imported into Python. It provides the functions \texttt{combinatoric}, \texttt{glynn}, \texttt{ryser}, and \texttt{opt}, which each take a single argument: the 2-dimensional NumPy array (\texttt{numpy.ndarray}) whose permanent is to be computed.
\begin{minted}{pycon}
>>> import permanent
>>> matrix = np.array([[1, 2, 3], [4, 5, 6], [7, 8, 9]])
>>> permanent.opt(matrix)
450
\end{minted}

\section{Benchmarks}

The absolute and relative performance of the computation of the permanent will, obviously, be influenced by the input matrix characteristics, including size, sparsity, and data type. As such, we assess the performance of the most efficient algorithms on a variety of input matrices and aggregate the results in order to determine benchmarks defining which algorithm to use based on the features of a given input matrix. 

We focus on two main criterion for the assessment, namely, relative execution time and precision. By nature the execution time itself depends on the hardware used, so we report relative execution time--the ratio of each algorithm's execution time as compared to the fastest algorithm. 

To obtain the benchmarks reported herein we used a sequential implementation on an Apple M1 Pro CPU (ARM-architecture) with an \mintinline{c++}{-O3} level of compiler optimization. We also compiled and optimized the library on a high-performance cluster equipped with Intel Xeon Gold 6448Y processors (x86\_64 architecture). By compiling and tuning the library on these diverse architectures, we ensure our solution is robust and versatile. The M1 optimization ensures excellent performance on modern, energy-efficient ARM-based systems that are increasingly common for personal computing environments. Meanwhile, the supercluster optimization guarantees that the library can scale to meet the demands of high-performance computing scenarios.

\begin{figure}[htbp]
  \centering
  \includegraphics[width=0.95 \linewidth]{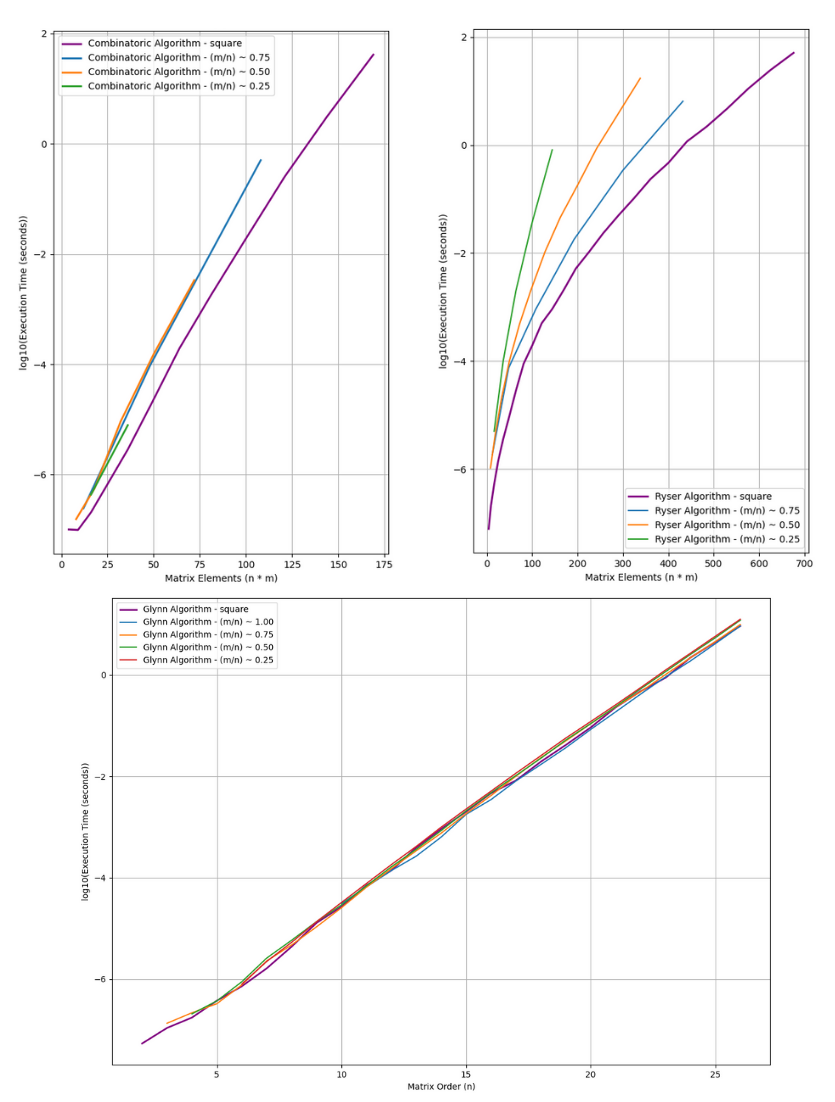}
  \caption{A comparison of the algorithm's execution time for evaluating the permanent of real-valued matrices with elements randomly selected from the interval [-1, 1]. The varying matrix aspect ratios are shown by the coloured lines for each algorithm. For the nai\"ve (combinatoric) algorithm, matrices with more than 14 columns were not considered because the time ($> 15$ minutes) exceeded that for all other algorithms ($< 1$ second) by three orders of magnitude.}
  \label{fig:plot_time}
\end{figure}

To test the performance of the algorithms, we generated random integer matrices (every element was randomly chosen as either zero or one) and random real matrices (elements were selected from the interval $[-1,1]$). The relative performance of the algorithms was similar in all three cases. The combinatoric algorithm is very expensive, and is actually somewhat slower for rectangular matrices. The Ryser algorithm is faster, and effectively exploits the reduced number of matrix elements in rectangular matrices. While the Glynn algorithm is fastest for near-square matrices, because it adds rows to rectangular matrices to make them square, it does not benefit from rectangularity; see \ref{fig:plot_time}.

The accuracy of the algorithms were assessed using the logarithm of the relative error divided by machine precision,
\begin{equation}\label{eq:digits_lost}
    d = \log_{10} \frac{|\text{evaluated} - \text{true}|}{|\text{true}|} - \log_{10}{\texttt{macheps}}.
\end{equation}
This represents the number of digits of precision that are lost during the calculation. As this formula requires the true value of the permanent, we assess the methods' accuracy using matrices where the true value of the permanent is known analytically. For this reason, the following matrices were used when assessing the precision:
\begin{itemize}
    \item All entries are ones (the permanent is $n!$ (or $\frac{n!}{(n-m)!}$)).
    \item The identity matrix, $\delta_{ij}$ (the permanent is $1$). For rectangular matrices, the extra columns are filled with zeros. 
    \item A Cauchy matrix where we randomly sample the vectors \textbf{x} and \textbf{y} from a uniform distribution in the range $[0.25, 0.75]$ and $[-0.75,-0.25]$ respectively, and then construct the matrix $\textbf{C}$ using the formula $C_{ij}=\frac{1}{x_i+y_j}$. The permanent is given by Eq. (\ref{eq:caucy_permanent}). The choice of parameters for the Cauchy matrix was chosen to avoid having very small denominators (very large elements) in the Cauchy matrix; this limits the growth of the size of the permanents. In addition, we repeat the sampling process for the vectors \textbf{x} and \textbf{y} 100,000 times and select the matrix with the lowest condition number. This provided us with reasonable condition numbers, though for very large matrices the Cauchy permanent can still be very large, leading to a loss of precision.
\end{itemize}

\begin{figure}[htbp]
  \centering
  \includegraphics[width=\linewidth]{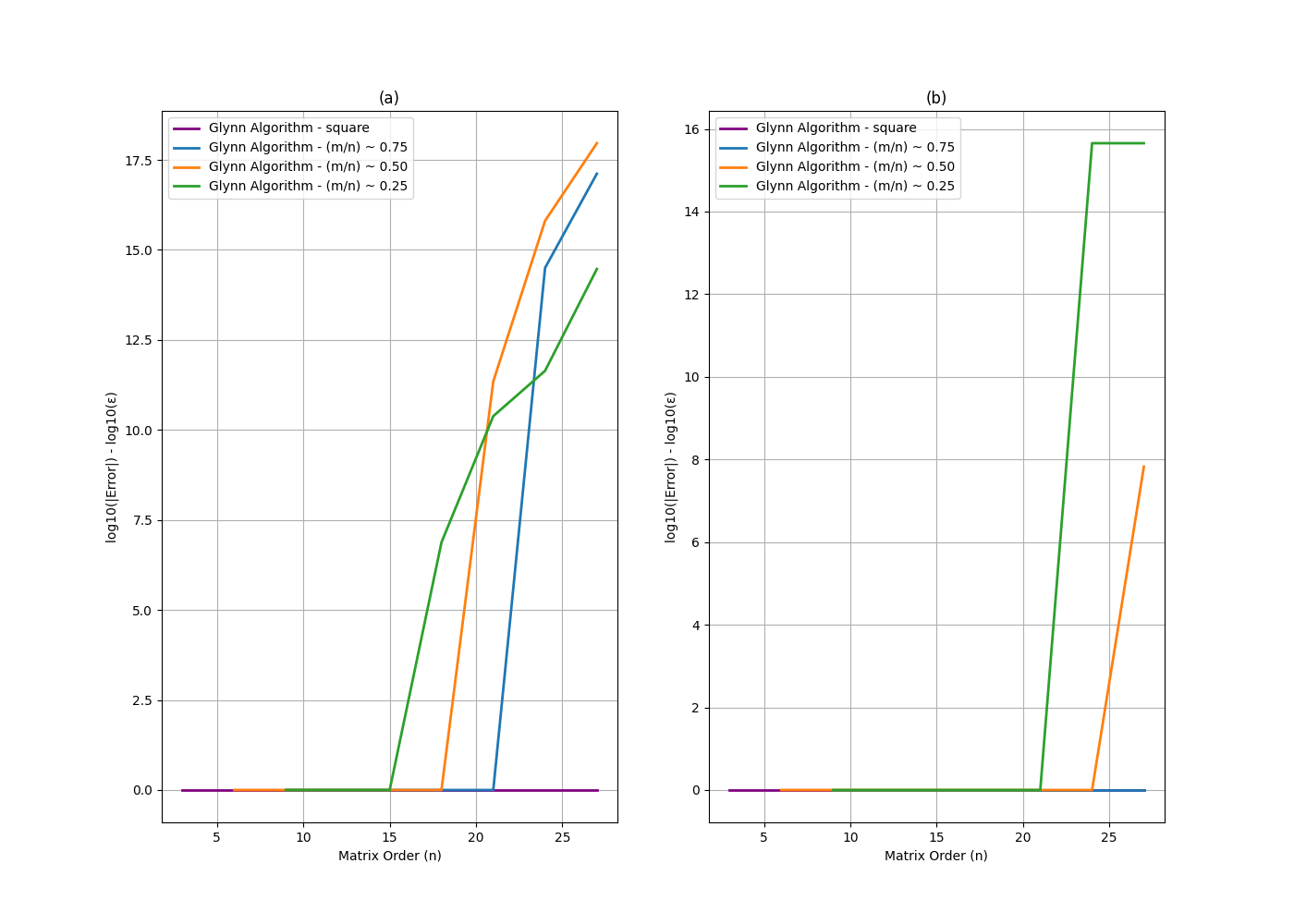}
  \caption{Accuracy of the Glynn algorithm for $n$-by-$n$ identity matrix with double (a) and integer (b) type. The error is assessed using Eq. (\ref{eq:digits_lost}). No other algorithms are displayed as the accuracy remained stable within the testing period.}
  \label{fig:glynn_tolerance}
\end{figure}

\begin{figure}[htbp]
  \centering
  \includegraphics[width=\linewidth]{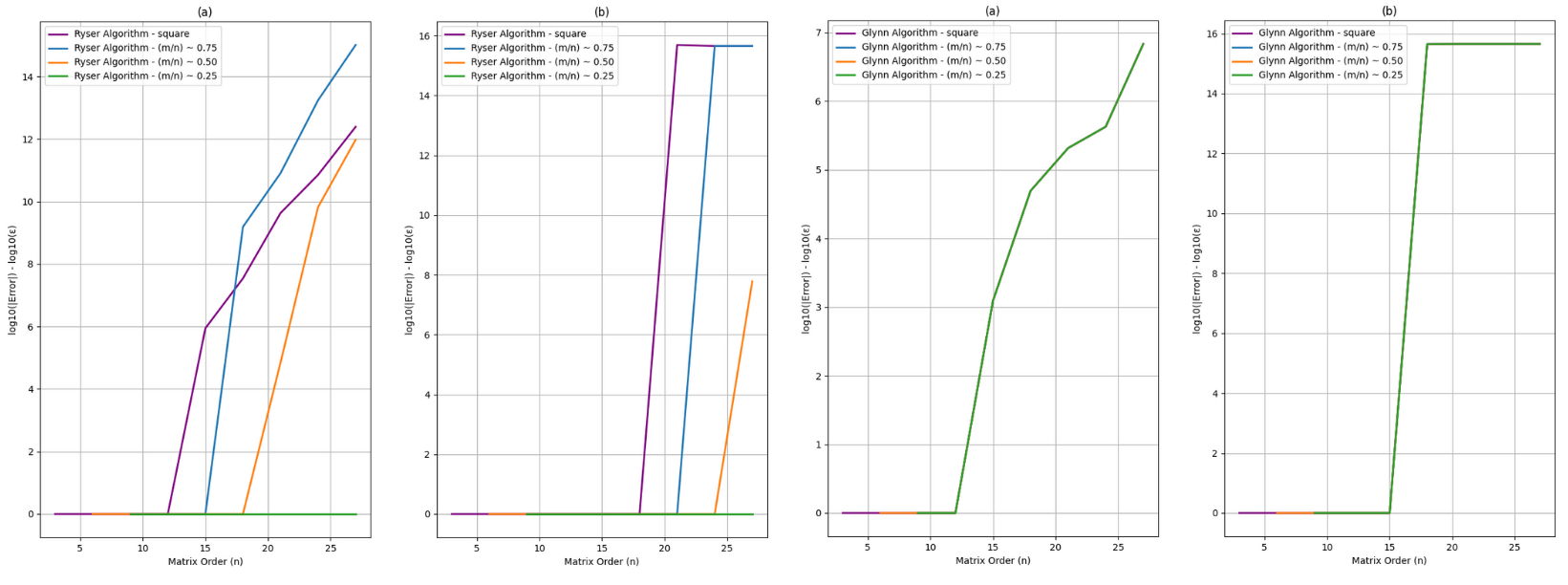}
  \caption{Accuracy of the Ryser and Glynn algorithm for $n$-by-$n$ ones matrix (every element is one) with double (a) and integer (b) type. The error is assessed using Eq. (\ref{eq:digits_lost}). Overflow occurs at the same moment for all matrix aspect ratios for the Glynn algorithm.}
  \label{fig:tolerance_ones}
\end{figure}

\begin{figure}[htbp]
  \centering
  \includegraphics[width=\linewidth]{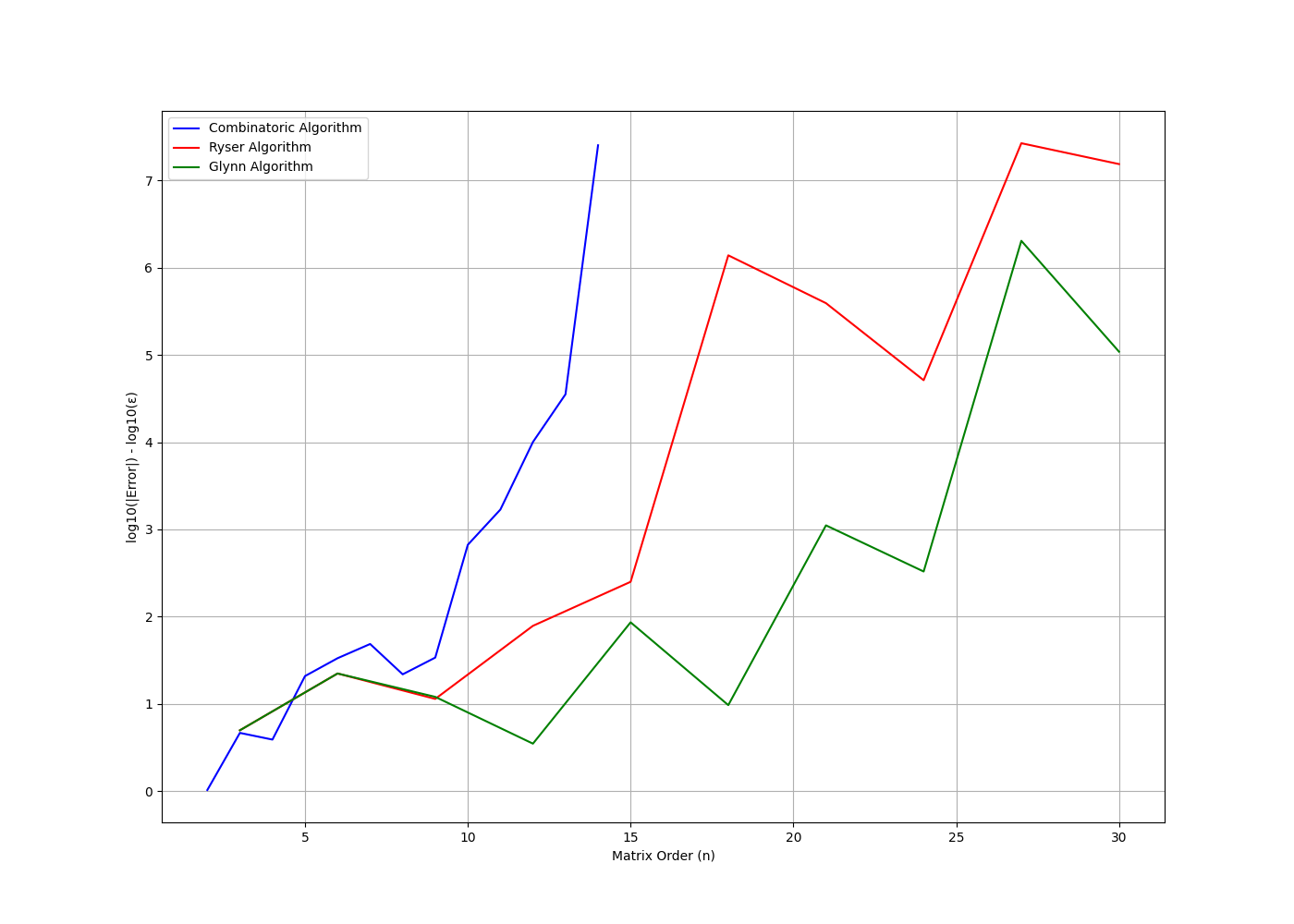}
  \caption{Accuracy of the algorithms for $n$-by-$n$ double-precision Cauchy matrices. The error is assessed using Eq. (\ref{eq:digits_lost}). The performance of the different algorithms is displayed by their corresponding colour, and we use the sampled matrix with the lowest condition number.}
  \label{fig:cauchy_tolerance}
\end{figure}

The precision of the algorithms for (square) Cauchy matrices is displayed in Figure \ref{fig:cauchy_tolerance}. (We only used the na\"ive (combinatic algorithm) for $n \le 14$ because it is extremely inefficient for larger matrices, and also seemingly less accurate than the other algorithms.) None of the algorithms gives good precision for larger matrices, probably because the size of the permanent and the intermediates used to compute it increase rapidly with matrix size, inducing accumulation of floating point roundoff errors.

Although we do not expect the (relative) execution time of the algorithms to depend on the data type of the matrix elements, the precision can change due to overflow and round-off errors. When assessing the performance of the algorithms for the (1) ones and (2) identity matrices, we also vary the input type between (3) double and (4) integer. For the (1) ones matrix, round-off errors accumulate rapidly, as would be expected from the combinatoric growth of the value of the permanent. The Ryser algorithm achieves somewhat better precision. Matrices with (4) integer elements maintain full precision for longer, but lose precision abruptly once the permanent gets too large. For the (2) identity matrix, the Ryser algorithm never loses precision, but the Glynn algorithm loses precision for non-square matrices. This is unsurprising since the rectangular matrices are padded by ill-conditioned rows of 1's in our algorithm for rectangular Glynn matrices.

\section{Summary}

The \texttt{matrix-permanent} library efficiently computes the permanent of general, rectangular, matrices. Users have the flexibility to either leverage the pre-computed default tuning, which selects the most efficient algorithm based on our own assessments of computational performance or, alternatively, to obtain user-specific tuning that optimizes the library's performance for their particular system architecture and/or use case. Furthermore, \texttt{matrix-permanent} is provided as both a \cpluspluslogo{} library and a Python package, thereby combining out-of-the-box utility with customizable optimization options, so that users can adapt the library to their needs. 

This development closes a significant gap in the open-source software community. Previously, the ability to flexibly integrate various efficient algorithms for computing matrix permanents was not available, despite the fact that the efficiency of these algorithms is heavily influenced by the order, shape, and characteristics of the input matrix. Given the importance of evaluating matrix permanents for a wide range of applications, from quantum mechanics, to machine learning, to quantum computing, we believe \texttt{matrix-permanent} has broad utility for the scientific community. Indeed, \texttt{matrix-permanent} has already been integrated into \texttt{PyCI}, where it is used to support emerging methods for modelling of quantum many-boson and many-fermion systems.

\section{Authors contributions}
Using the CRediT system: Conceptualization: CM, MR \& PWA; Data curation: CM; Formal analysis: PWA; Investigation: PWA; Methodology: CM, MR \& PWA. 
Project administration: PWA; Resources: PWA; Software: CM, MR \& PWA; Supervision: PWA; Validation: CM \& MR; 
Visualization: CM; Writing -- original draft: CM, MR \& PWA; Writing -- review and editing: CM, MR \& PWA.

\section{Acknowledgment}
The authors acknowledge the support of the QC-Devs Team (). PWA acknowledges the Natural Sciences and Engineering Research Council (NSERC) of Canada, the Canada Research Chairs, and the Digital Research Alliance of Canada (DRAC) for financial and computational support. 

\newpage
\appendix

\section{Algorithm base implementations} \label{app:algorithms}
This appendix contains the base algorithmic implementations of the algorithms described in the main text. These implementations form the foundation of the released software package \href{https://github.com/theochem/matrix-permanent}{matrix-permanent}, and are included here to provide a precise reference for reproducibility. The algorithms are presented in simplified \cpluspluslogo{} style pseudocode, focusing on the core logic rather than low-level optimizations. 

\subsection{Ryser’s Algorithm} \label{app:ryser_algorithms}
The inclusion–exclusion formulation of Ryser’s method (see Section~\ref{ryser_algo}) leads to efficient evaluation of the permanent with scaling $\mathcal{O}(2^n \cdot n)$.  
For completeness, we provide the pseudocode implementations here.

\begin{algorithm}[H]
\caption{Ryser's Algorithm for Square Matrices}\label{app:alg:ryser_square}
\begin{algorithmic}
\Require Square matrix $A$ of size $m \times m$
\Ensure Permanent of matrix $A$
\State Initialize $\text{out} \gets 0$
\State Set $c \gets 2^m$ \Comment{Total number of subsets}
\For{$k = 0$ to $c - 1$} \Comment{Iterate over all subsets}
    \State Initialize $\text{rowsumprod} \gets 1$
    \For{$i = 0$ to $m - 1$} \Comment{For each row}
        \State Initialize $\text{rowsum} \gets 0$
        \For{$j = 0$ to $m - 1$} \Comment{For each column}
            \If{$k \land 2^j \neq 0$} \Comment{If column $j$ is in subset $k$}
                \State $\text{rowsum} \gets \text{rowsum} + A_{i,j}$
            \EndIf
        \EndFor
        \State $\text{rowsumprod} \gets \text{rowsumprod} \times \text{rowsum}$
    \EndFor
    \State $\text{sign} \gets (-1)^{\text{popcount}(k)}$ \Comment{Alternating sign based on subset size}
    \State $\text{out} \gets \text{out} + \text{rowsumprod} \times \text{sign}$
\EndFor
\State $\text{final\_sign} \gets (-1)^m$ if $m$ is odd, else $1$
\State \Return $\text{out} \times \text{final\_sign}$
\end{algorithmic}
\end{algorithm}

\begin{algorithm}[H]
\caption{Ryser's Algorithm for Rectangular Matrices}\label{app:alg:ryser_rectangular}
\begin{algorithmic}
\Require Matrix $A$ of size $m \times n$ where $m \leq n$
\Ensure Permanent of matrix $A$
\State Initialize $\text{sign} \gets 1$
\State Initialize $\text{out} \gets 0$
\For{$k = 0$ to $m - 1$} \Comment{Iterate over subset sizes}
    \State Generate all combinations $C(n, m-k)$ of size $(m-k)$ from $n$ columns
    \State Compute $\text{bin} \gets \binom{n-m+k}{k}$ \Comment{Binomial coefficient}
    \State Initialize $\text{permsum} \gets 0$
    \For{each combination $\text{comb}$ in $C(n, m-k)$} \Comment{For each column subset}
        \State Initialize $\text{colprod} \gets 1$
        \For{$i = 0$ to $m - 1$} \Comment{For each row}
            \State Initialize $\text{matsum} \gets 0$
            \For{$j = 0$ to $(m-k) - 1$} \Comment{Sum over selected columns}
                \State $\text{matsum} \gets \text{matsum} + A_{i,\text{comb}[j]}$
            \EndFor
            \State $\text{colprod} \gets \text{colprod} \times \text{matsum}$
        \EndFor
        \State $\text{permsum} \gets \text{permsum} + \text{colprod} \times \text{sign} \times \text{bin}$
    \EndFor
    \State $\text{out} \gets \text{out} + \text{permsum}$
    \State $\text{sign} \gets \text{sign} \times (-1)$ \Comment{Alternate sign for next iteration}
\EndFor
\State \Return $\text{out}$
\end{algorithmic}
\end{algorithm}

\subsection{Glynn’s Algorithm} \label{app:glynn_algorithms}
The invariant-theory formulation by Glynn (see Section~\ref{glynn_algo}) provides an alternative expression for the permanent, also scaling as $\mathcal{O}(2^n \cdot n)$.  
The pseudocode is given below.

\begin{algorithm}[H]
\caption{Glynn's Algorithm for Square Matrices}\label{app:alg:glynn_square}
\begin{algorithmic}
\Require Square matrix $A$ of size $m \times m$
\Ensure Permanent of matrix $A$
\State Initialize $\delta[i] \gets 1$ for $i = 0, 1, \ldots, m-1$ \Comment{Sign array}
\State Initialize $\text{perm}[i] \gets i$ for $i = 0, 1, \ldots, m-1$ \Comment{Permutation array}
\State \Comment{Handle first permutation}
\State Initialize $\text{out} \gets 1$
\For{$j = 0$ to $m - 1$} \Comment{For each column}
    \State $\text{sum} \gets 0$
    \For{$i = 0$ to $m - 1$} \Comment{Compute weighted column sum}
        \State $\text{sum} \gets \text{sum} + A_{i,j} \times \delta[i]$
    \EndFor
    \State $\text{out} \gets \text{out} \times \text{sum}$
\EndFor
\State \Comment{Iterate through remaining permutations}
\State Initialize $\text{bound} \gets m - 1$, $\text{pos} \gets 0$, $\text{sign} \gets 1$
\While{$\text{pos} \neq \text{bound}$}
    \State $\text{sign} \gets \text{sign} \times (-1)$ \Comment{Update sign}
    \State $\delta[\text{bound} - \text{pos}] \gets \delta[\text{bound} - \text{pos}] \times (-1)$ \Comment{Flip delta}
    \State Initialize $\text{prod} \gets 1$
    \For{$j = 0$ to $m - 1$} \Comment{Compute term for current permutation}
        \State $\text{sum} \gets 0$
        \For{$i = 0$ to $m - 1$}
            \State $\text{sum} \gets \text{sum} + A_{i,j} \times \delta[i]$
        \EndFor
        \State $\text{prod} \gets \text{prod} \times \text{sum}$
    \EndFor
    \State $\text{out} \gets \text{out} + \text{sign} \times \text{prod}$
    \State \Comment{Generate next permutation}
    \State $\text{perm}[0] \gets 0$
    \State $\text{perm}[\text{pos}] \gets \text{perm}[\text{pos} + 1]$
    \State $\text{pos} \gets \text{pos} + 1$
    \State $\text{perm}[\text{pos}] \gets \text{pos}$
    \State $\text{pos} \gets \text{perm}[0]$
\EndWhile
\State \Return $\text{out} / 2^{\text{bound}}$ \Comment{Divide by normalization factor}
\end{algorithmic}
\end{algorithm}

\clearpage

\begin{breakablealgorithm}
    \caption{Glynn's Algorithm for Rectangular Matrices}\label{app:alg:glynn_rectangular}
    \begin{algorithmic}
    \Require Matrix $A$ of size $m \times n$ where $m \leq n$
    \Ensure Permanent of matrix $A$
    \State Initialize $\delta[i] \gets 1$ for $i = 0, 1, \ldots, n-1$ \Comment{Extended sign array}
    \State Initialize $\text{perm}[i] \gets i$ for $i = 0, 1, \ldots, n-1$ \Comment{Permutation array}
    \State \Comment{Handle first permutation}
    \State Initialize $\text{out} \gets 1$
    \For{$j = 0$ to $n - 1$} \Comment{For each column}
        \State $\text{sum} \gets 0$
        \For{$i = 0$ to $m - 1$} \Comment{Sum over matrix rows}
            \State $\text{sum} \gets \text{sum} + A_{i,j} \times \delta[i]$
        \EndFor
        \For{$k = m$ to $n - 1$} \Comment{Sum over extended delta entries}
            \State $\text{sum} \gets \text{sum} + \delta[k]$
        \EndFor
        \State $\text{out} \gets \text{out} \times \text{sum}$
    \EndFor
    \State \Comment{Iterate through remaining permutations}
    \State Initialize $\text{bound} \gets n - 1$, $\text{pos} \gets 0$, $\text{sign} \gets 1$
    \While{$\text{pos} \neq \text{bound}$}
        \State $\text{sign} \gets \text{sign} \times (-1)$ \Comment{Update sign}
        \State $\delta[\text{bound} - \text{pos}] \gets \delta[\text{bound} - \text{pos}] \times (-1)$ \Comment{Flip delta}
        \State Initialize $\text{prod} \gets 1$
        \For{$j = 0$ to $n - 1$} \Comment{Compute term for current permutation}
            \State $\text{sum} \gets 0$
            \For{$i = 0$ to $m - 1$}
                \State $\text{sum} \gets \text{sum} + A_{i,j} \times \delta[i]$
            \EndFor
            \For{$k = m$ to $n - 1$}
                \State $\text{sum} \gets \text{sum} + \delta[k]$
            \EndFor
            \State $\text{prod} \gets \text{prod} \times \text{sum}$
        \EndFor
        \State $\text{out} \gets \text{out} + \text{sign} \times \text{prod}$
        \State \Comment{Generate next permutation}
        \State $\text{perm}[0] \gets 0$
        \State $\text{perm}[\text{pos}] \gets \text{perm}[\text{pos} + 1]$
        \State $\text{pos} \gets \text{pos} + 1$
        \State $\text{perm}[\text{pos}] \gets \text{pos}$
        \State $\text{pos} \gets \text{perm}[0]$
    \EndWhile
    \State \Return $\frac{\text{out}}{2^{\text{bound}} \times (n-m)!}$ \Comment{Divide by normalization factors}
    \end{algorithmic}
\end{breakablealgorithm}

\section{Optimized Algorithm Selection} \label{app:opt_algorithm}
In Section~\ref{sec:using_cpp_lib}, we introduced the \texttt{opt} variant of the library interface, which selects between the available permanent algorithms (combinatorial, Glynn, and Ryser) depending on matrix size and aspect ratio. This adaptive selection is controlled by a set of tunable parameters. For completeness, we provide the pseudocode implementation here.

\begin{algorithm}[H]
\caption{Optimized Algorithm Selection}\label{app:alg:opt_permanent}
\begin{algorithmic}
\Require Matrix $A$ of size $m \times n$ where $m \leq n$
\Require Tuning parameters $\{p_1, p_2, p_3, p_4, p_5, p_6, p_7, p_8\}$
\Ensure Permanent of matrix $A$ using optimal algorithm
\State Compute aspect ratio $r \gets m/n$ \Comment{For square matrices: $r = 1$}
\If{$n \leq p_8$} \Comment{Small matrix regime ($n \leq 13$)}
    \If{$m = n$ and $n \leq p_4$} \Comment{Very small square matrices only}
        \State \Return \textsc{Combinatorial}$(A)$ \Comment{Brute force enumeration}
    \Else \Comment{Small-medium matrices}
        \State Evaluate hyperplane: $h_1 \gets p_1 \cdot r + p_2 \cdot n + p_3$
        \If{$h_1 > 0$} \Comment{Above first decision boundary}
            \State \Return \textsc{Combinatorial}$(A)$ \Comment{Square or rectangular}
        \Else \Comment{Below first decision boundary}
            \State \Return \textsc{Glynn}$(A)$ \Comment{Square or rectangular variant}
        \EndIf
    \EndIf
\Else \Comment{Large matrix regime ($n > 13$)}
    \State Evaluate hyperplane: $h_2 \gets p_5 \cdot r + p_6 \cdot n + p_7$
    \If{$h_2 > 0$} \Comment{Above second decision boundary}
        \State \Return \textsc{Glynn}$(A)$ \Comment{Square or rectangular variant}
    \Else \Comment{Below second decision boundary}
        \State \Return \textsc{Ryser}$(A)$ \Comment{Square or rectangular variant}
    \EndIf
\EndIf
\end{algorithmic}
\end{algorithm}

\newpage
\bibliography{references}

\end{document}